\documentclass[prd,twocolumn,showpacs,showkeys,preprintnumbers,floatfix,nofootinbib,superscriptaddress]{revtex4-1}
\usepackage{amsfonts} 
\usepackage{amssymb} 
\usepackage{amsmath} 
\usepackage{graphicx} 
\usepackage{subfigure} 
\usepackage{array} 
\usepackage{dcolumn} 
\usepackage{bm} 
\usepackage{latexsym} 
\usepackage{longtable} 
\usepackage{hyperref} 
\usepackage{bbold}
\usepackage{color}
\usepackage{multirow}
\usepackage{rotating}
\usepackage{comment}

\graphicspath{{./figs/}}

\definecolor{green}{rgb}{0.1, 0.8, 0.1}

\begin{document}


\title{Flavor diagonal tensor charges of the nucleon from 2+1+1 flavor lattice QCD}
\author{Rajan Gupta}
\email{rajan@lanl.gov}
\affiliation{Los Alamos National Laboratory, Theoretical Division T-2, Los Alamos, NM 87545, USA}

\author{Boram Yoon}
\email{boram@lanl.gov}
\affiliation{Los Alamos National Laboratory, CCS Division CCS-7, Los Alamos, NM 87545, USA}

\author{Tanmoy Bhattacharya}
\email{tanmoy@lanl.gov}
\affiliation{Los Alamos National Laboratory, Theoretical Division T-2, Los Alamos, NM 87545, USA}

\author{Vincenzo Cirigliano}
\email{cirigliano@lanl.gov}
\affiliation{Los Alamos National Laboratory, Theoretical Division T-2, Los Alamos, NM 87545, USA}

\author{Yong-Chull Jang}
\email{ypj@bnl.gov}
\affiliation{Brookhaven National Laboratory, Upton, NY 87545, USA}

\author{Huey-Wen~Lin}
\email{hwlin@pa.msu.edu}
\affiliation{Department of Physics and Astronomy, Michigan State University, East Lansing, MI, 48824, USA}
\affiliation{Department of Computational Mathematics,  Science and Engineering, Michigan State University, East Lansing, MI 48824, USA}

\collaboration{PNDME}
\preprint{LA-UR-18-28007}
\preprint{MSUHEP-18-016}
\date{\today}
\begin{abstract}
We present state-of-the-art results for the matrix elements of flavor
diagonal tensor operators within the nucleon state. The calculation of
the dominant connected contribution is done using eleven ensembles of
gauge configurations generated by the MILC Collaboration using the
highly improved staggered quark (HISQ) action with 2+1+1 dynamical flavors. The calculation of the
disconnected contributions is done using seven (six) ensembles for the
strange (light) quarks.  These high-statistics simulations allowed us
to address various systematic uncertainties. A simultaneous fit in the
lattice spacing and the light-quark mass is used to extract the tensor
charges in the continuum limit and at $M_\pi=135$~MeV. Results for the
proton in the $\overline{MS}$ scheme at 2~GeV are: $g_T^u =
0.784(28)(10)$, $g_T^d = -0.204(11)(10)$ and $g_T^s = -0.0027(16)$.
Implications of these results for constraining the quark electric
dipole moments and their contributions to the neutron electric dipole
moment are discussed.
\end{abstract}
\maketitle
%
%
%
%
\section{Introduction}
\label{sec:into}

High precision calculations of the matrix elements of flavor diagonal
quark bilinear operators, $\overline{q} \Gamma q $ where $\Gamma $ is
one of the sixteen Dirac matrices, within the nucleon state provide a
quantitative understanding of a number of properties of nucleons and
their interactions with electrically neutral probes. In this paper, we
present results for the tensor charges, $g_T^u$, $g_T^d$ and $g_T^s$,
that give the contribution of the electric dipole moment (EDM) of
these quark flavors to the EDM of the nucleon. They are defined as the
nucleon matrix elements of the renormalized tensor operator, $Z_T
\overline{q} \sigma^{\mu\nu} q$ with $\sigma^{\mu\nu} =
i[\gamma_\mu,\gamma_\nu]/2$, $Z_T$ the renormalization constant and
$q$ the bare quark field:
\begin{align}
\langle N(p’,s’) |  Z_T \bar{q} \sigma_{\mu \nu} q | N (p,s) \rangle =  
g_T^{q}   \ \bar{u}_N (p’,s’) \sigma_{\mu \nu} u_N (p,s)  \,.
\label{eq:gTdef}
\end{align}
Experimentally, they can be extracted from semi-inclusive
deep-inelastic scattering (SIDIS)
data~\cite{Lin:2017stx,Radici:2018iag,Ye:2016prn}.  These tensor
charges also provide the hadronic input to the weakly interacting massive 
particle (WIMP)-nucleus cross
section in dark matter models that generate tensor quark-WIMP
operators~\cite{Bishara:2017pfq}.

New high-statistics data for both the connected and disconnected
contributions to the tensor charges allow us to control the various
systematic uncertainties and perform a chiral-continuum fit to obtain
physical results.  The light-quark disconnected contributions, which were
neglected in our previous
works~\cite{Bhattacharya:2015wna,Bhattacharya:2015esa}, are $O(0.01)$, 
nevertheless the data are precise enough to allow 
extrapolation to the continuum limit and $M_\pi =135$~MeV.  We also 
report a signal in the still smaller $g_T^s$, whose contribution to the
neutron EDM can be enhanced versus $g_T^u$ by $ m_s/m_u \approx 40$ in
models in which the chirality flip is provided by the Standard Model
Yukawa couplings.

\section{Lattice Methodology}
\label{sec:Methodology}

All the calculations were done on ensembles with 2+1+1 flavors of highly improved 
staggered quarks (HISQ) 
fermions~\cite{Follana:2006rc} generated by the MILC
Collaboration~\cite{Bazavov:2012xda}.  In order to calculate the
matrix elements of flavor diagonal operators, one needs to evaluate
the contribution of both the ``connected'' and ``disconnected''
diagrams.  The lattice methodology and our strategy for the calculation and
analysis of the two-point and connected three-point functions using Wilson
clover fermions on the HISQ ensembles has been described in
Refs.~\cite{Bhattacharya:2015wna,Bhattacharya:2016zcn,Yoon:2016dij,Rajan:2017lxk}
and for the disconnected contribution in
Refs.~\cite{Bhattacharya:2015wna,Lin:2018obj}.

The details of the calculation and analysis of the connected
contributions on eleven ensembles covering the range 0.15--0.06~fm in
the lattice spacing, $M_\pi =$ 135--320~MeV in the pion mass, and
$M_\pi L =$ 3.3--5.5 in the lattice size have been presented in
Ref.~\cite{Gupta:2018qil} and readers are referred to it. With these
high-statistics data ($O(10^5)$ measurements on $O(1000)$
configurations on each of the 11 ensembles), a
chiral-continuum-finite-volume fit in the three variables $a$,
$M_\pi^2$ and $M_\pi L$ was performed to control the systematic uncertainties
due to lattice discretization, dependence on the quark mass and finite
lattice size.  The final results, in the $\overline{MS}$ scheme at
2~GeV, for the connected contribution to the proton, are reproduced from
Ref.~\cite{Gupta:2018qil}:\looseness-1
\begin{align}
g_T^{u-d} &=  0.989(32)   \qquad g_T^{u+d}|_{\rm conn} &=  0.590(25) \,, \nonumber \\
g_T^{u}|_{\rm conn}   &=  0.790(27)   \qquad g_T^{d}|_{\rm conn}   &=  -0.198(10) \,.
\label{eq:gTconn}
\end{align}

In this paper, we focus on the analysis of the disconnected
contributions using seven ensembles with simulation parameters given
in Table~\ref{tab:ens}.  By combining these with the connected
contributions given in Eq.~\eqref{eq:gTconn}, we obtain final results
for the flavor diagonal tensor charges.

\begin{table}[tbp]    
\begin{center}
\renewcommand{\arraystretch}{1.2} 
\begin{ruledtabular}
\begin{tabular}{l|cc|cc|c}
Ensemble      & $N_\text{conf}^l$ & $N_\text{src}^l$ & $N_\text{conf}^s$ & $N_\text{src}^s$  &  $N_{\rm LP}/N_{\rm HP}$  \\
\hline
$a15m310 $    & 1917              & 2000             & 1919              &  2000             &  50  \\
\hline
$a12m310 $    & 1013              & 5000             & 1013              &  1500             &  30   \\
$a12m220 $    &  958              & 11000            & 958               &  4000             &  30   \\
\hline                                             
$a09m310 $    & 1081              & 4000             & 1081              &  2000             &  30   \\
$a09m220 $    & 712               & 8000             & 847               &  10000            &  30/50   \\
$a09m130 $    &                   &                  & 877               &  10000            &  50   \\
\hline                                             
$a06m310 $    & 830               & 4000             & 200+340           &  5000+10000       &  50   \\
\end{tabular}
\end{ruledtabular}
\caption{The number of configurations analyzed for the light ($N_\text{conf}^l$) and
  strange ($N_\text{conf}^s$) quarks, the corresponding number of random sources 
  ($N_\text{src}$) sampled, and the ratio $N_{\rm LP}/N_{\rm HP}$ of low to high precision
  measurements made to estimate the disconnected quark loop contribution on each configuration.  
    }
\label{tab:ens}
\end{center}
\end{table}

\section{Controlling Excited-State Contamination (ESC)}
\label{sec:ESC}

\begin{figure*}[hptb]
\begin{center}                                               
  \subfigure{
    \includegraphics[height=1.0in,trim={0.20cm 0.70cm 0 0.06cm},clip]{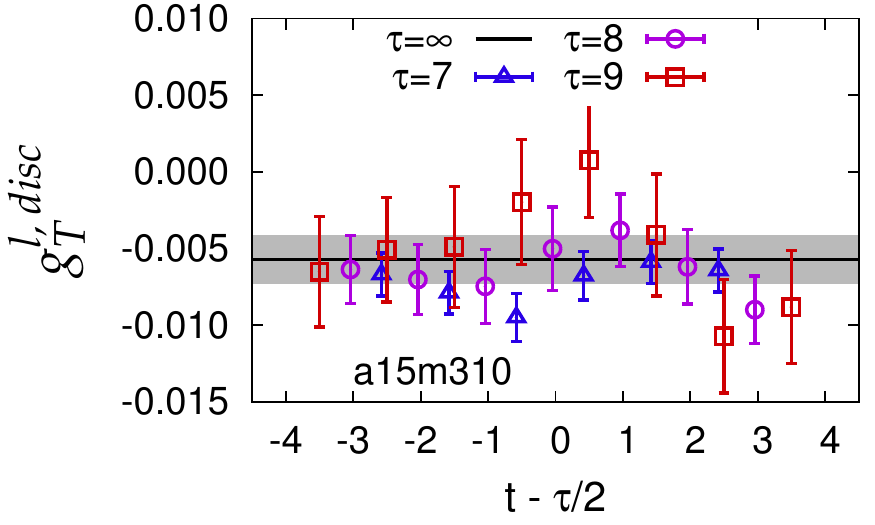}
    \includegraphics[height=1.0in,trim={1.20cm 0.70cm 0 0.06cm},clip]{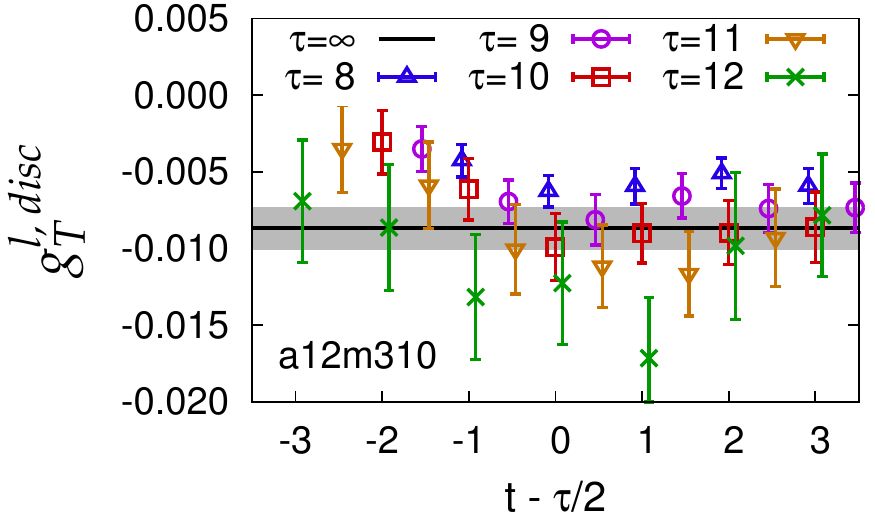}
    \includegraphics[height=1.0in,trim={1.20cm 0.70cm 0 0.06cm},clip]{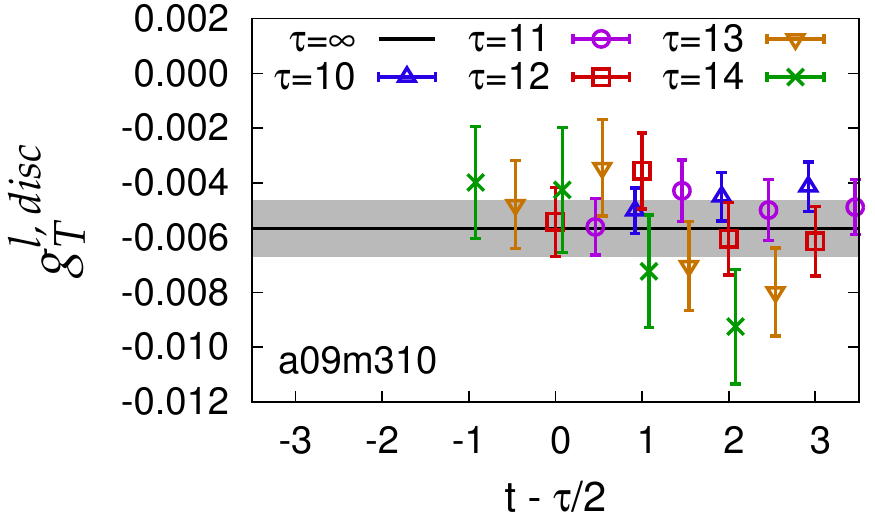}
    \includegraphics[height=1.0in,trim={1.20cm 0.70cm 0 0.06cm},clip]{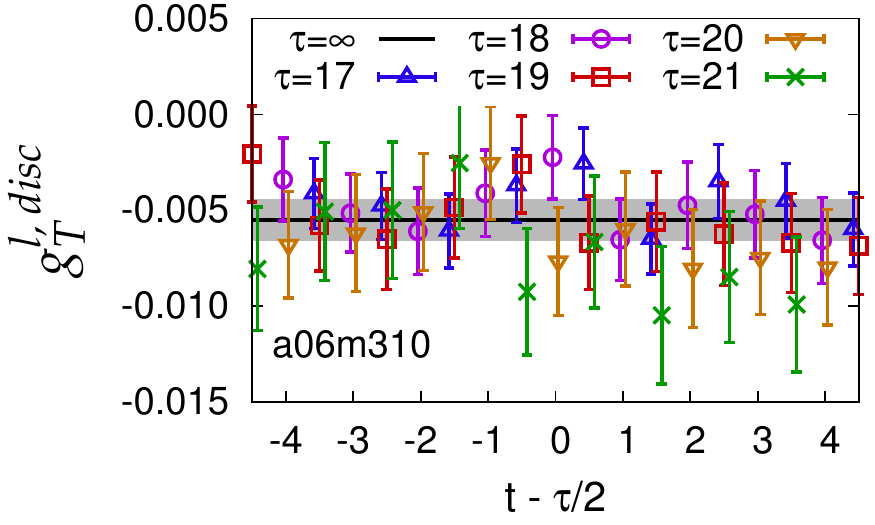}
  }\\
  \subfigure{
    \includegraphics[height=1.2in,trim={0.20cm 0.10cm 0 0.06cm},clip]{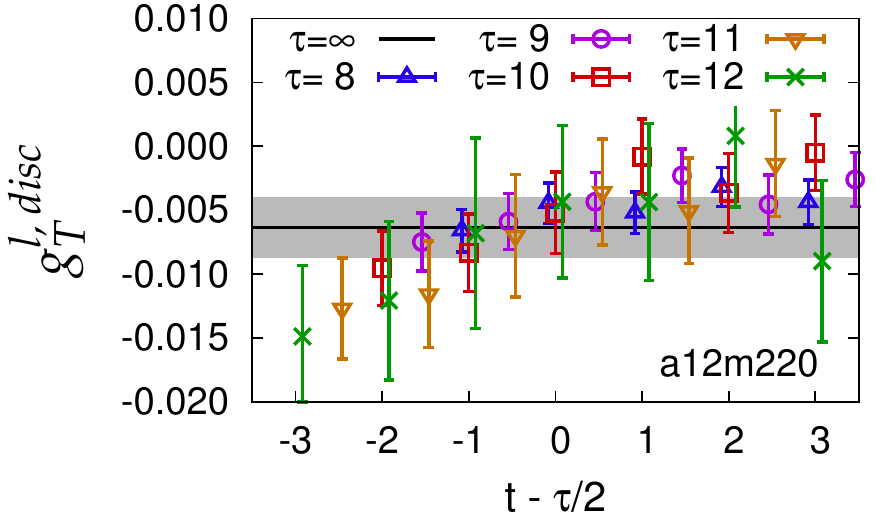}
    \includegraphics[height=1.2in,trim={1.20cm 0.10cm 0 0.06cm},clip]{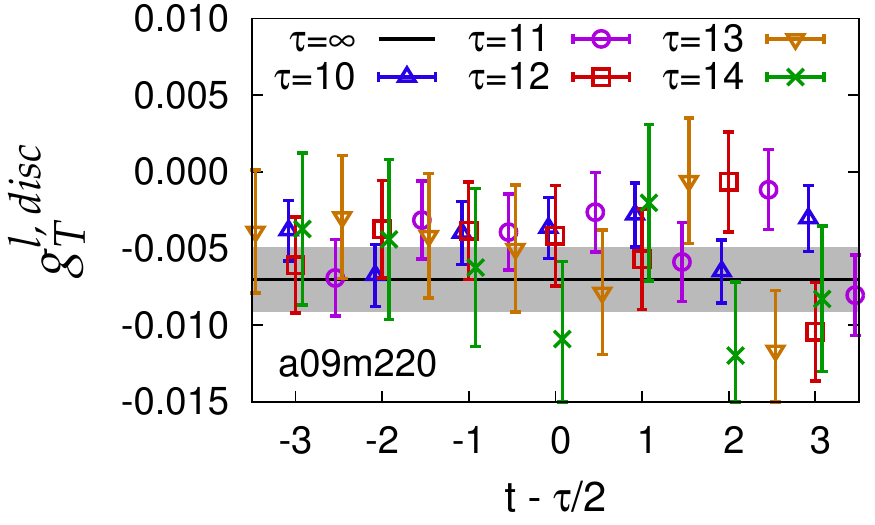}
  }
  \subfigure{
    \includegraphics[height=1.0in,trim={0.20cm 0.70cm 0 0.06cm},clip]{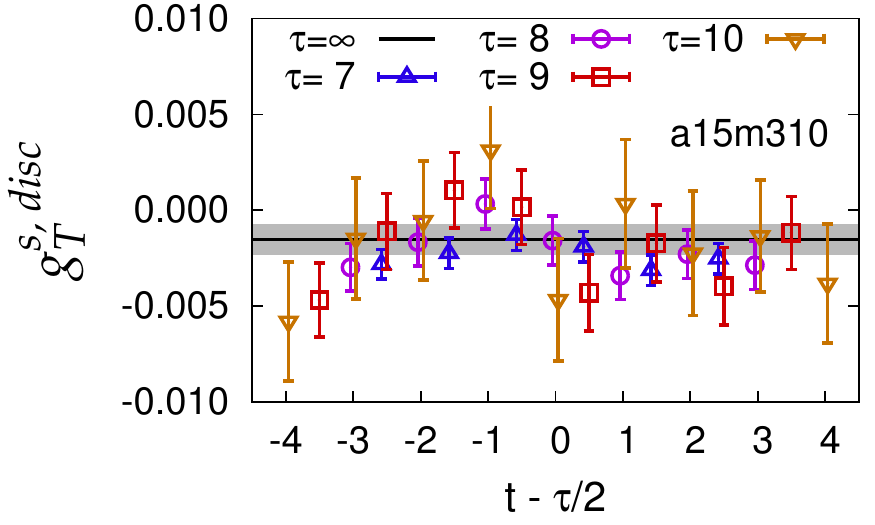}
    \includegraphics[height=1.0in,trim={1.20cm 0.70cm 0 0.06cm},clip]{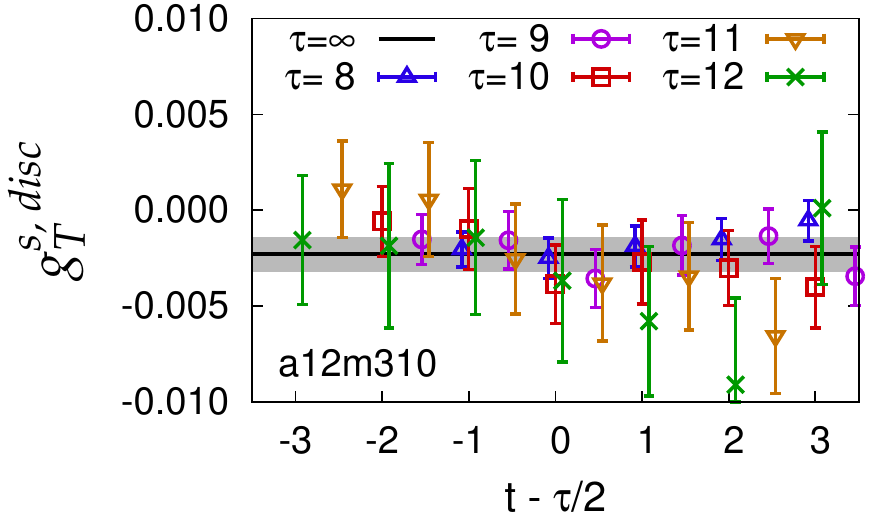}
    \includegraphics[height=1.0in,trim={1.20cm 0.70cm 0 0.06cm},clip]{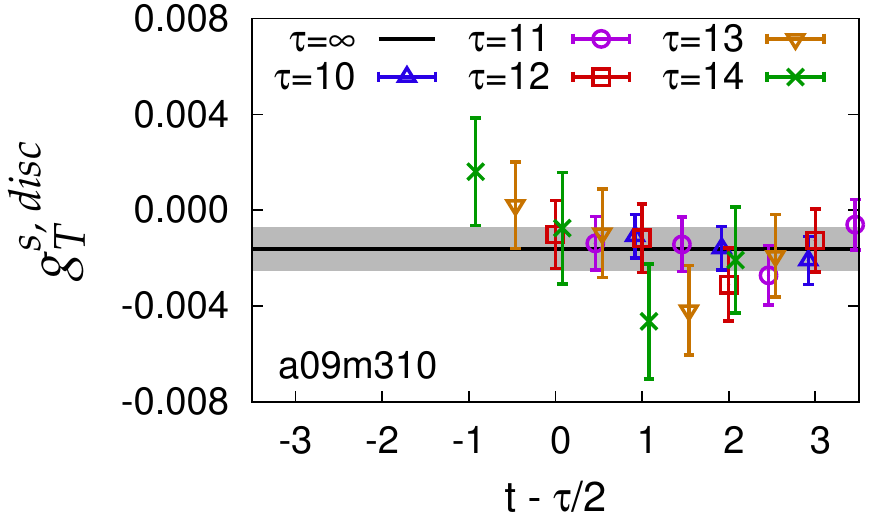}
    \includegraphics[height=1.0in,trim={1.20cm 0.70cm 0 0.06cm},clip]{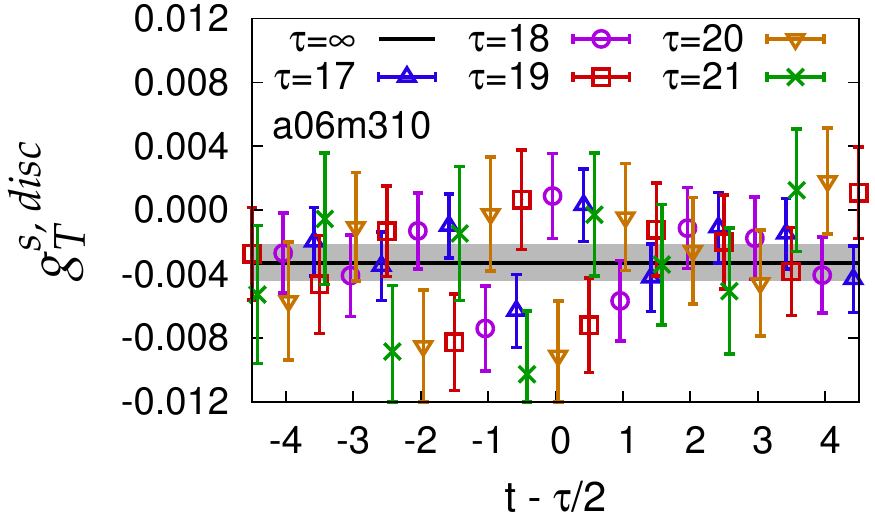}
  }\\
  \subfigure{
    \includegraphics[height=1.2in,trim={0.20cm 0.10cm 0 0.06cm},clip]{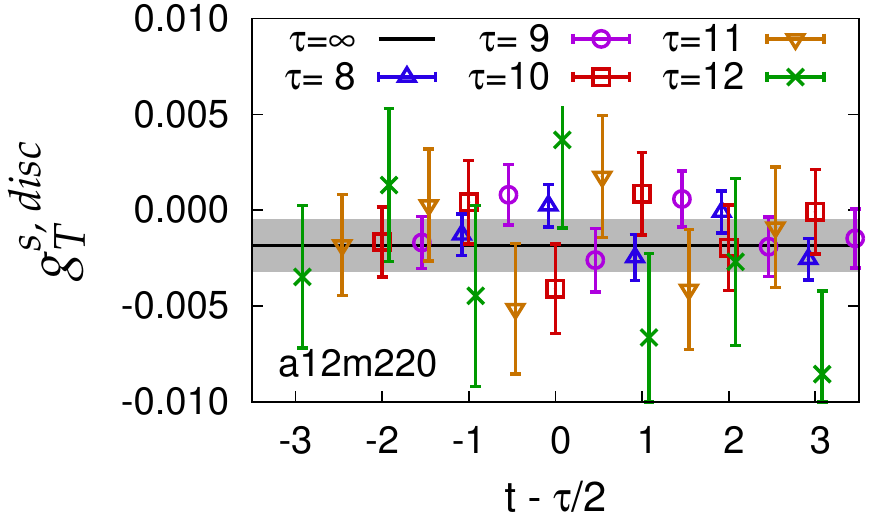}
    \includegraphics[height=1.2in,trim={1.20cm 0.10cm 0 0.06cm},clip]{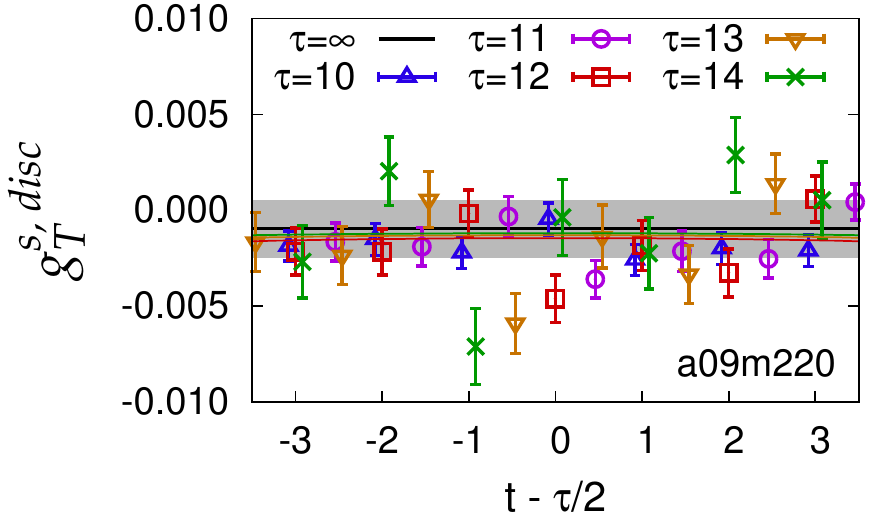}
    \includegraphics[height=1.2in,trim={1.20cm 0.10cm 0 0.06cm},clip]{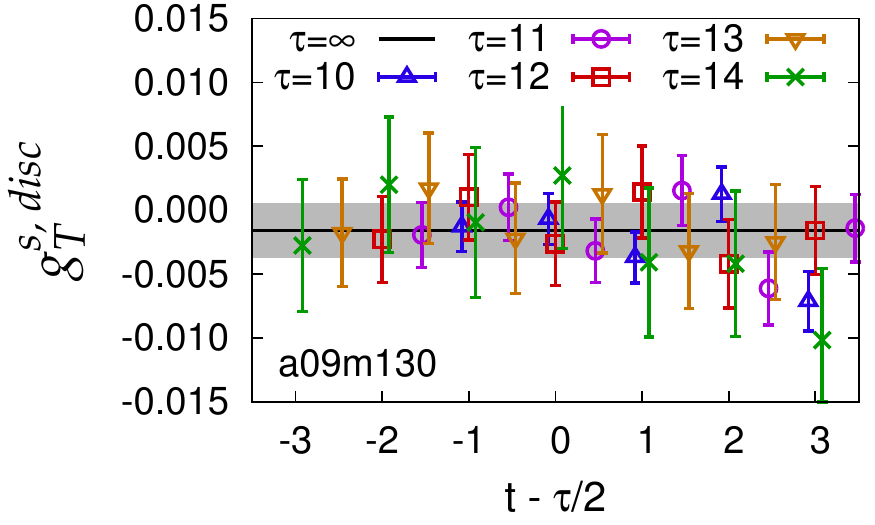}
  }
  \end{center}  
\vspace{-0.4cm}                                                                            
\caption{The data for the unrenormalized disconnected contributions of
  the light $g_T^{l, {\text disc}}$ (top two rows) and strange
  $g_T^{s, {\text disc}}$ (bottom two rows) quarks.  The ground state
  estimate is given by the solid black line within the gray band.  It
  is obtained as the average over data at multiple $t$ and $\tau$
  since no significant variation versus them is observed.
  \label{fig:gTESC}}
\end{figure*}

The first step in the analysis is to understand and remove the excited
state contamination (ESC) in the disconnected contribution. A number
of features stand out in the data shown in Fig.~\ref{fig:gTESC}.
First, for a given value of the source-sink separation $\tau$, the
data are much more noisy compared to the corresponding connected
contribution analyzed in Ref.~\cite{Gupta:2018qil}. Second, within
statistical uncertainties, there is no discernible variation with
$\tau$.  In fact, the data at the various values of $\tau$ overlap for
both the light, $g_{T}^{l}$, and the strange, $g_{T}^{s}$ quark
contributions: i.e., no ESC is apparent in either.  Lastly, the
magnitude, in most cases, is smaller than $0.01$, which is smaller
than the statistical uncertainty in the connected contribution. 
Possible residual ESC is expected to be even smaller. The bottom line is, for
the estimate on each ensemble, we take a simple average over the
multiple $t$ and $\tau$ data shown in Fig.~\ref{fig:gTESC}. These results for
the bare charges, $g_{T}^{l}$ and $g_{T}^{s}$, are compiled together
in Table.~\ref{tab:resultsgT}. \looseness-1

In Refs.~\cite{Gupta:2018qil,Lin:2018obj}, we raised the need for
evaluating the uncertainty due to analyzing the connected and
disconnected contributions separately to remove the ESC using the QCD
spectral decomposition. For the tensor charges, this uncertainty is
expected to be negligible for two reasons: The disconnected data show
no evidence for ESC and we take the average over the various $\tau$
values, i.e., no fits using the spectral decomposition are
made. Second, the magnitude of the contributions, $< 0.01$, is 
smaller than the combined statistical errors.  So, we assume that any
residual uncertainty due to performing separate analyses will be even
smaller. \looseness-1

%
%
\begin{table*}   
\centering
\begin{ruledtabular}
\begin{tabular}{c|cc|cc|cc|cc}
Ensemble & $a$     & $M_\pi$ & $g_T^{l}|_{\rm bare}$ & $g_T^{s}|_{\rm bare}$  &  $g_T^{l}|_{R1}$ & $g_T^{s}|_{R1}$ &  $g_T^{l}|_{R2}$ & $g_T^{s}|_{R2}$   \\ 
ID       & (fm)    & (MeV)   &                       &                        &                  &                 &                  &                   \\ 
\hline                                                                                                                                    
a15m310 & 0.151(2) & 320(5)  & $-$0.0057(16) & $-$0.0015(8)   &  $-$0.0054(15) & $-$0.0014(8)  &  $-$0.0054(15) & $-$0.0014(8)    \\
\hline                                                                                                                                    
a12m310 & 0.121(1) & 310(3)  & $-$0.0086(14) & $-$0.0023(9)   &  $-$0.0081(14) & $-$0.0021(8)  &  $-$0.0084(14) & $-$0.0022(9)    \\
a12m220 & 0.118(1) & 228(2)  & $-$0.0063(24) & $-$0.0018(14)  &  $-$0.0059(23) & $-$0.0017(13) &  $-$0.0061(23) & $-$0.0017(14)   \\
\hline                                                                                                                                    
a09m310 & 0.089(1) & 313(3)  & $-$0.0057(10) & $-$0.0016(9)   &  $-$0.0056(10) & $-$0.0016(9)  &  $-$0.0058(10) & $-$0.0017(9)    \\
a09m220 & 0.087(1) & 226(2)  & $-$0.0070(21) & $-$0.0016(9)   &  $-$0.0069(21) & $-$0.0016(9)  &  $-$0.0071(21) & $-$0.0016(9)    \\
a09m130 & 0.087(1) & 138(1)  &               & $-$0.0016(21)  &                & $-$0.0016(21) &                & $-$0.0016(21)   \\
\hline                                                                                                                                   
a06m310 & 0.058(1) & 320(2)  & $-$0.0055(11) & $-$0.0033(11)  &  $-$0.0057(12) & $-$0.0034(11) &  $-$0.0059(12) & $-$0.0035(12)   \\
\end{tabular}
\end{ruledtabular}
\caption{The values of $a$ and $M_\pi$ for the seven
  ensembles are given in columns 2 and 3.  Results for the
  unrenormalized disconnected light and strange quark contributions,
  $g_{T}^{l,s}$, are given in columns four and five.  They are
  obtained using a simple average over the data shown in
  Fig.~\protect\ref{fig:gTESC} since no significant ESC is
  evident. Columns 2 and 3 give the lattice spacing of the HISQ
  ensembles and the valence $M_\pi$, as described in
  Ref.~\cite{Gupta:2018qil}.  In columns 6--9, we give the
  renormalized charges $g_T^{l,s}|_{R1}$ and $g_T^{l,s}|_{R2}$ defined
  in Eq.~\eqref{eq:renorm}.  The isovector renormalization constant
  $Z_T^{\rm isovector}$ is used in all cases as discussed in the
  text. \looseness-1 }
\label{tab:resultsgT}
\end{table*}

\section{Renormalization of the operators}
\label{sec:renorm}

Flavor diagonal light-quark operators, $\overline{q} \Gamma q$, can be
written as a sum over isovector ($u-d$) and isoscalar ($u+d$)
combinations which renormalize differently---isovector with $Z^{\rm
  isovector}$ and isoscalar with $Z^{\rm isoscalar}$. The difference
between $Z^{\rm isovector}$ and $Z^{\rm isoscalar}$ for quark bilinear
operators starts, in general, at two loops in perturbation theory. For
the tensor operator, the two-loop terms are zero because the spin trace
vanishes in the clover, HISQ, and thus clover-on-HISQ
formulations~\cite{Constantinou:2016ieh}.  Also, for the twisted mass
action, nonperturbative calculations show $Z_T^{\rm isovector} =
Z_T^{\rm isoscalar}$ to within a
percent~\cite{Alexandrou:2017qyt,Alexandrou:2017oeh}. We have not
calculated $Z_T^{\rm isoscalar}$ nonperturbatively for the
clover-on-HISQ formulation, which has additional $O(a)$ chiral breaking versus
the twisted mass action. In this work, we assume that the difference is
smaller than the statistical errors. The isovector renormalization
constants, $Z_T^{\rm isovector}$, calculated in the RI-sMOM scheme and
converted to the $\overline{MS}$ scheme at 2~GeV using two-loop
perturbation theory, are taken from Ref.~\cite{Gupta:2018qil}, and
used to renormalize the connected and disconnected contributions to
$g_T^{u}$, $g_T^{d}$ and $g_T^{s}$ in two ways:
\begin{align}
g_T^{l,s}|_{R1} &= g_T \times Z_T^{\rm isovector} \,, \nonumber \\
g_T^{l,s}|_{R2} &= \frac{g_T}{g_V^{u-d}} \times 
\frac{Z_T^{\rm isovector}}{ Z_V^{u-d}}   \,. 
\label{eq:renorm}
\end{align}
The conserved vector charge condition $g_V^{u-d} \times Z_V^{u-d}=1$
is implicit in the second definition. These two results for the
renormalized disconnected contributions on each ensemble are also
given in Table~\ref{tab:resultsgT}. They are extrapolated separately
to the continuum limit and $M_\pi=135$~MeV, and the extrapolated
results are given in Table~\ref{tab:resultsrenormgD}.

\section{The Continuum-Chiral Extrapolation}
\label{sec:CCFV}

\begin{figure*}[tbh]
\centering
  \subfigure{
    \includegraphics[height=1.18in,trim={0.1cm   0.10cm 0 0.1cm},clip]{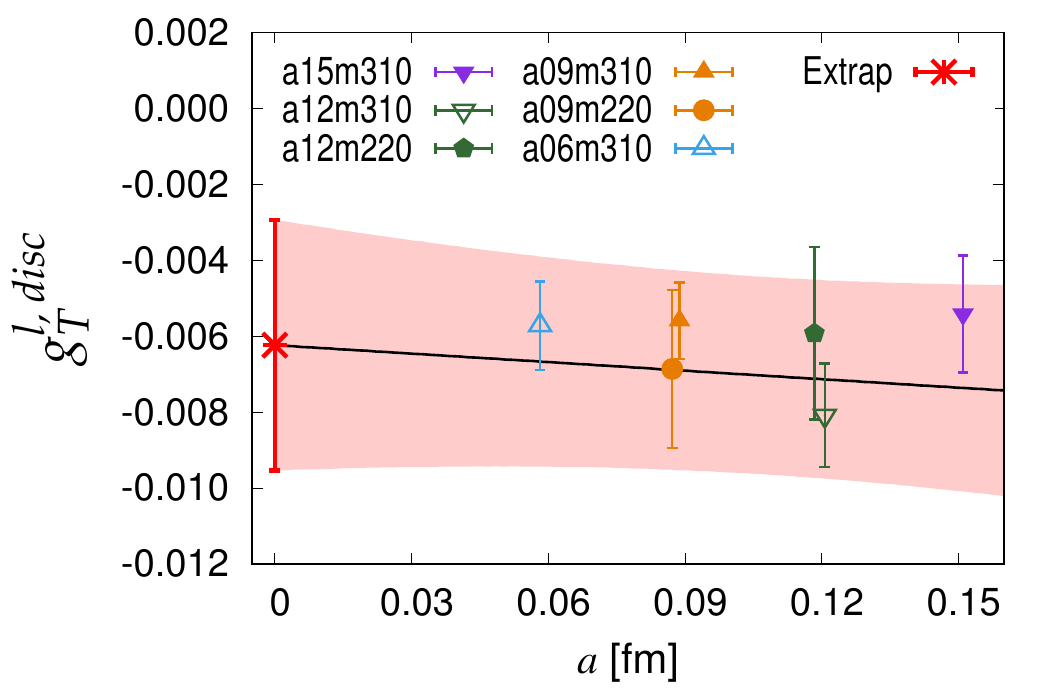} 
    \includegraphics[height=1.18in,trim={1.3cm   0.10cm 0 0.1cm},clip]{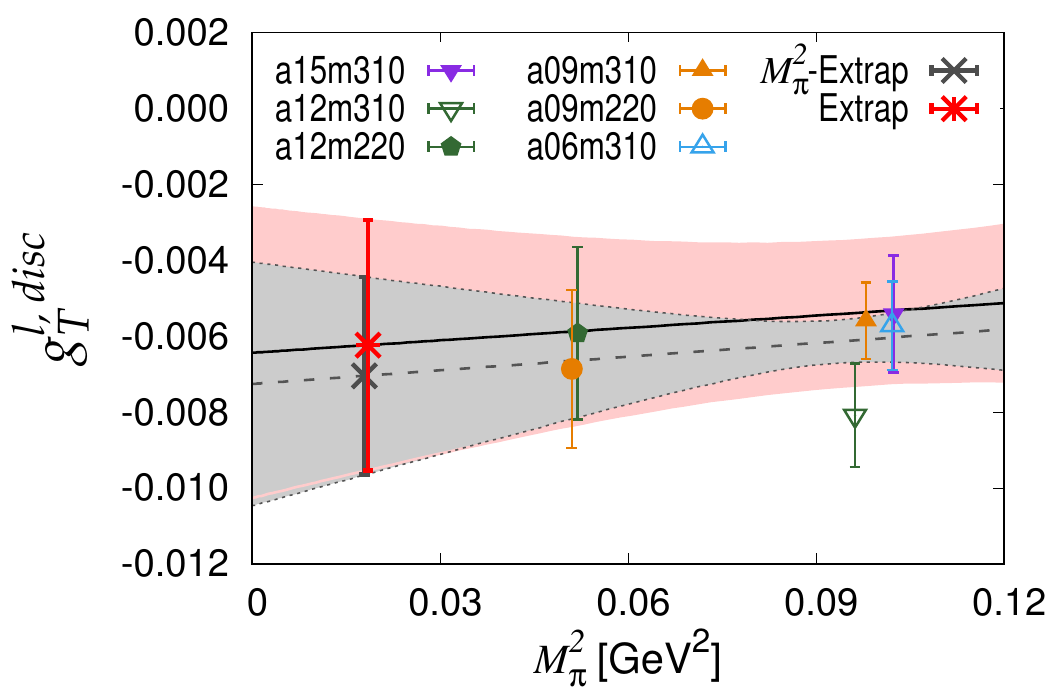}
    \includegraphics[height=1.18in,trim={0.1cm   0.10cm 0 0.1cm},clip]{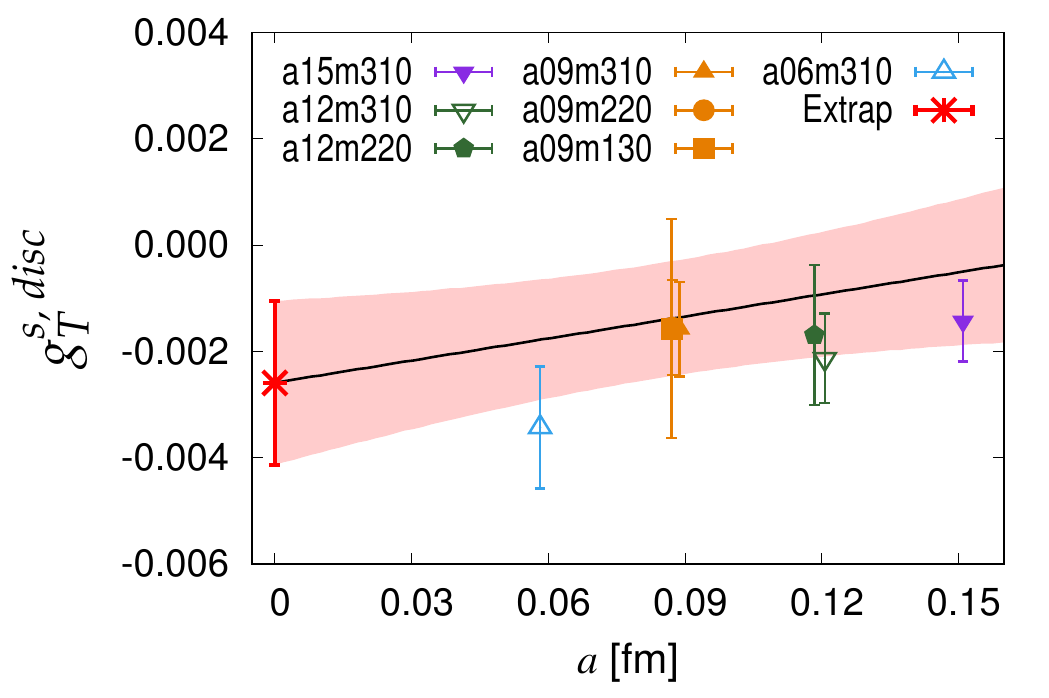} 
    \includegraphics[height=1.18in,trim={1.3cm   0.10cm 0 0.1cm},clip]{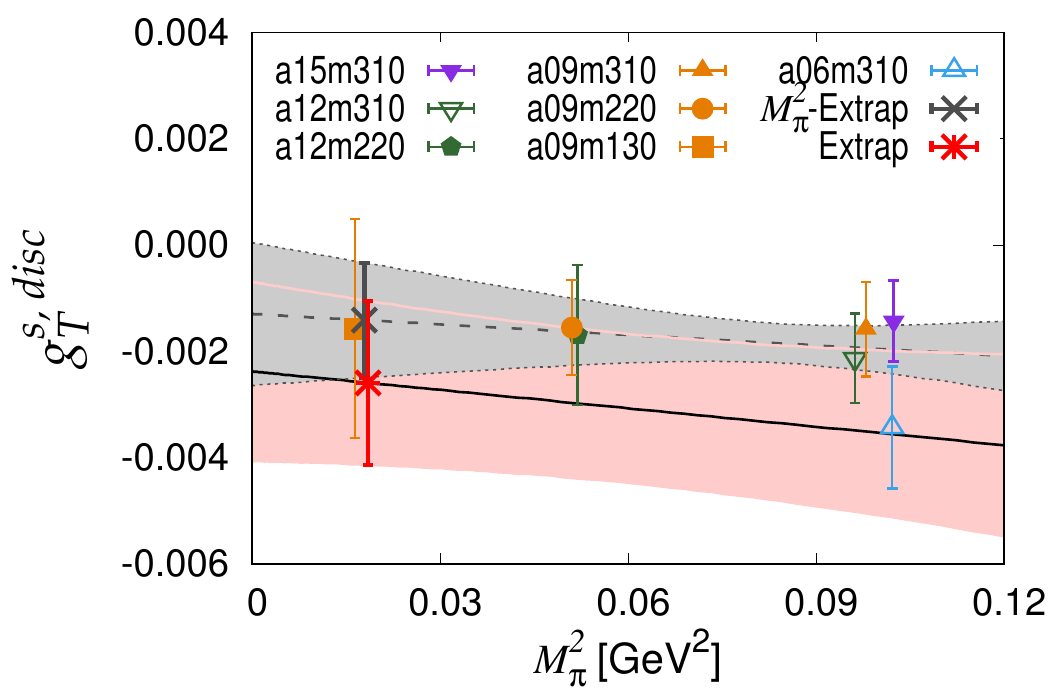} 
}
\caption{The data, in the $\overline{MS}$ scheme at 2~GeV, for the
  disconnected contribution $g_T^{l,{\rm disc}}|_{R2}$ (left two panels) and
  $g_T^{s,{\rm disc}}|_{R2}$ (right two panels) plotted versus $a$ and
  $M_\pi$. In each panel, the result at $a=0$ and $M_\pi=135$~MeV,
  obtained using Eq.~\protect\eqref{eq:extrapgT}, is shown by the red
  star. The pink band is the fit shown versus $a$ ($M_\pi$), with the
  other variable set to its physical value.  For comparison, the gray
  band between dotted lines shows a simpler linear fit versus only
  $M_\pi^2$, i.e., ignoring the dependence on $a$.
  \label{fig:gT-extrap}}
\end{figure*}

The last step in the analysis is to evaluate the results at
$M_{\pi^0}= 135$~MeV and in the continuum and infinite volume limits,
$a\rightarrow 0$ and $M_\pi L \to \infty$.  Over the limited range of
$M_\pi L $ spanned by our disconnected data, $3.9 < M_\pi L < 4.8$, 
finite-volume corrections were negligible in the connected contributions, 
as shown in Fig.~2 in Ref.~\cite{Gupta:2018qil}. We, therefore, assume possible
finite-volume corrections can be neglected in the disconnected contributions, and fit
the renormalized data given in Table~\ref{tab:resultsgT} keeping just
the leading correction terms in $a$ and $M_\pi$:
\begin{align}
  g_{T}^{l,s} (a,M_\pi,L) = c_1 + c_2a + c_3 M_\pi^2 +  \ldots \,, 
\label{eq:extrapgT} 
\end{align}
The data with the renormalization method $R2$ and the results of the
fits are shown in Fig.~\ref{fig:gT-extrap}.  The dependence of both
$g_T^{l}$ and $g_T^{s}$ on $M_\pi$ and $a$ is small and the
extrapolated value is consistent with an average over the six (seven)
points.  Given this consistency between the average values and the
results of the linear extrapolation using Eq.~\eqref{eq:extrapgT}, and
applying the Akaike Information Criteria~\protect\cite{1100705} (see
Table~\ref{tab:resultsrenormgD} for the $\chi^2/$DOF of the fits),
including additional higher order corrections to the chiral-continuum fit ansatz, 
Eq.~\eqref{eq:extrapgT}, is not warranted.

We consider the errors from the fit reasonable as they are
larger than those in most individual points and cover the total range
of variation in the points. Since the difference between the
extrapolated results given in Table~\ref{tab:resultsrenormgD} for the
two ways of doing the renormalization is much smaller than these 
errors, for the final value we take the average of the two as
summarized in Tables~\ref{tab:resultsrenormgD}
and~\ref{tab:resultsFINAL}. \looseness-1

In the connected contributions to $g_T^{u}$ and $g_T^{d}$, analyzed in
Ref.~\cite{Gupta:2018qil}, a systematic uncertainty of $0.01$ was
assessed to account for residual uncertainty in the
chiral-continuum-finite-volume fits made with only the leading order
corrections. This 0.01 is quoted as the second error in the final
results for $g_T^{u}$ and $g_T^{d}$ given in
Table~\ref{tab:resultsFINAL}.

\begin{table*}
\centering
\begin{ruledtabular}
\begin{tabular}{c|cc|cc|c|cc|cc|c}
                  &  \multicolumn{5}{c|}{light}                                       & \multicolumn{5}{c}{strange}                                                         \\
                  & $g_T^{l}|_{R1}$     & $\chi^2$/DOF & $g_T^{l}|_{R2}$     & $\chi^2$/DOF & $g_T^{l}$       & $g_T^{s}|_{R1}$     & $\chi^2$/DOF  & $g_T^{s}|_{R2}$     & $\chi^2$/DOF  & $g_T^{s}$        \\ 
\hline
$g_T^{\rm disc}$  & $-$0.0062(33) & 0.85            & $-$0.0066(33) & 0.91            & $-$0.0064(33) & $-$0.0026(15) & 0.31             & $-$0.0027(16) & 0.29             & $-$0.0027(16)  \\
\end{tabular}
\end{ruledtabular}
\caption{Results, in the $\overline{MS}$ scheme at 2~GeV, for the
  renormalized disconnected contributions to the proton's tensor
  charges were obtained in the limit $a = 0$ and $M_{\pi^0} = 135$~MeV
  by performing a chiral-continuum extrapolation using
  Eq.~\protect\eqref{eq:extrapgT}. The $\chi^2$/DOF of the two fits
  and the results (labeled $R1$ and $R2$) for the renormalized
  charges defined in Eq.~\protect\eqref{eq:renorm} are given along
  with the final results obtained by averaging $g_T^{l,s}|_{R1}$ and
  $g_T^{l,s}|_{R2}$ and taking the larger of the two errors.  }
\label{tab:resultsrenormgD}
\end{table*}

\begin{table}
\centering
\begin{ruledtabular}
\begin{tabular}{c|ccc}
              & $g_T^{u}$     & $g_T^{d}$    & $g_T^{s}$      \\ 
\hline
Connected     &    0.790(27)  & $-$0.198(10) &                \\
Disconnected  & $-$0.0064(33) & $-$0.0064(33)& $-$0.0027(16)  \\ 
\hline         
PNDME'18      &    0.784(28)(10)  & $-$0.204(11)(10) & $-$0.0027(16)  \\
\hline
ETMC'17~\protect\cite{Alexandrou:2017qyt}         &    0.782(21)  & $-$0.219(17) & $-0.00319(72)$  \\
PNDME'15~\protect\cite{Bhattacharya:2015wna}      &    0.774(66)  & $-$0.233(28) & $ $0.008(9)  \\
\end{tabular}
\end{ruledtabular}
\caption{Final results, in the $\overline{MS}$ scheme at 2~GeV, for
  the individual connected and disconnected contributions to the
  flavor diagonal tensor charges and their sum, labeled PNDME'18 in
  the third row.  The fourth row gives the ETMC
  results~\protect\cite{Alexandrou:2017qyt} for comparison. These were
  obtained from a single physical mass ensemble at $a=0.0938(4)$ and
  $M_\pi=130.5(4)$~MeV , i.e., without a continuum extrapolation and
  at small $M_\pi L = 2.98$. Comparing PNDME'18 and PNDME'15
  results~\protect\cite{Bhattacharya:2015wna}, highlights the
  improvements realized with higher statistics and more ensembles, in particular, we now 
  present results for light-quark disconnected contributions. \looseness-1 }
\label{tab:resultsFINAL}
\end{table}

\begin{figure*}[tbh]
\begin{flushleft}                                               
  \subfigure{
    \includegraphics[width=.55\textwidth]{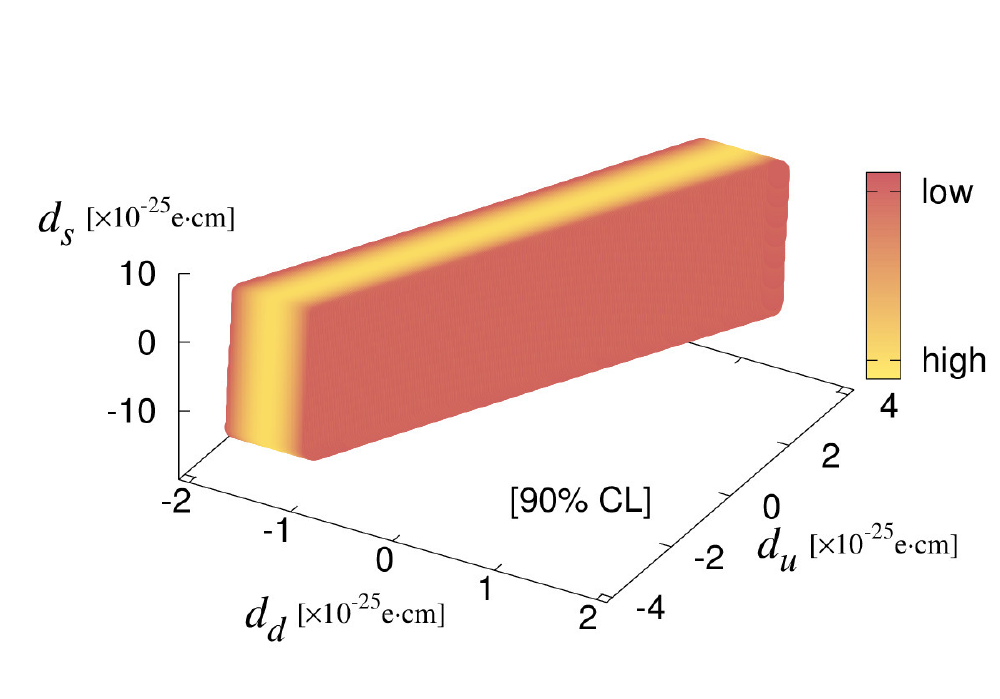} 
    \includegraphics[width=.48\textwidth]{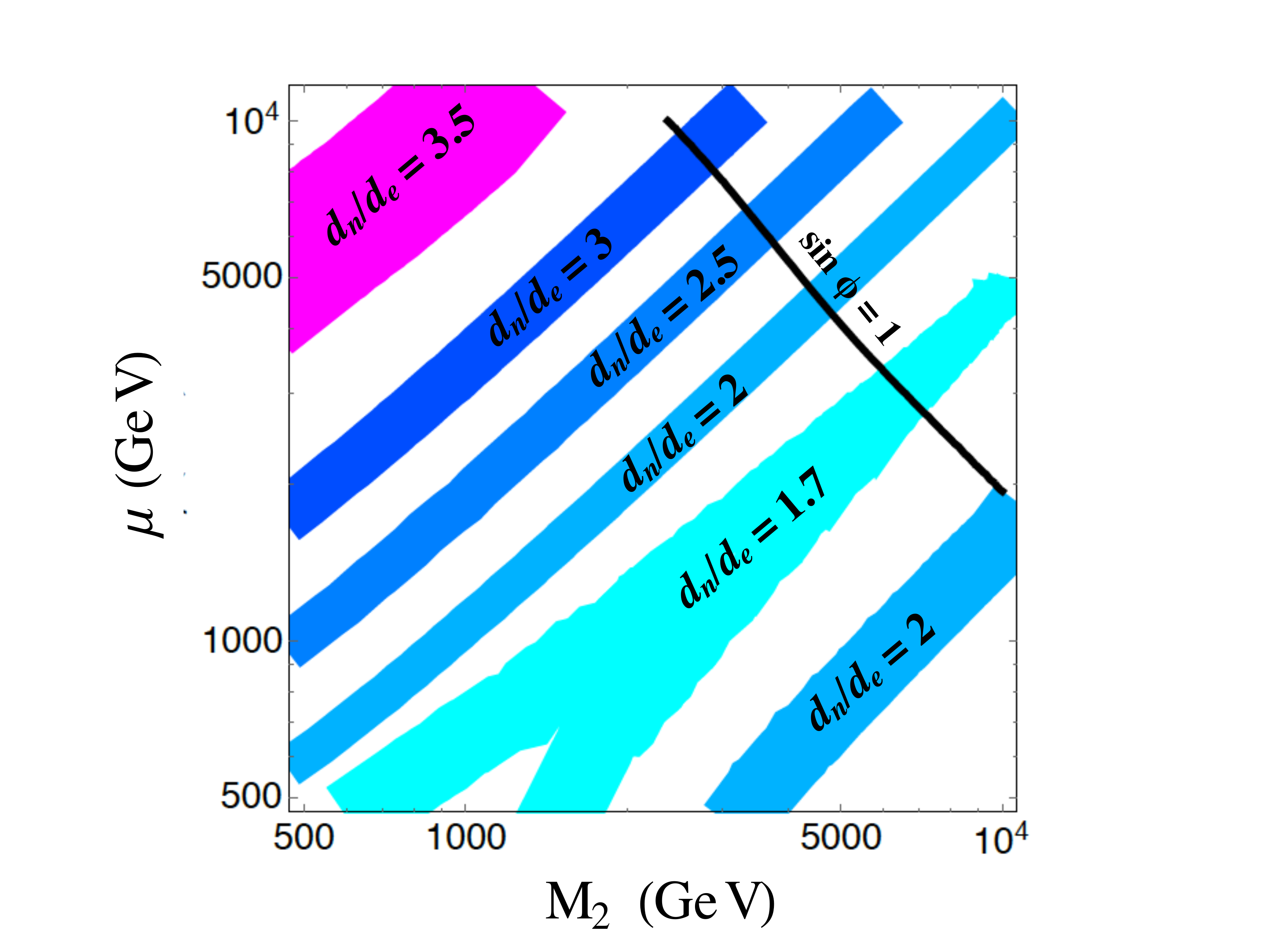} 
  }
 \end{flushleft}                                                      
\caption{(Left) Constraints on the BSM couplings of the $CP$-violating
  quark EDM operator using the current experimental bound on the nEDM
  ($2.9\times 10^{-26}\ e$ cm~\protect\cite{Baker:2006ts}) and
  assuming that only these couplings contribute. The strongest
  constraint is a strip in $d_u$ and $d_d$, i.e., representing the
  thickness of the slab, with high (low) corresponding to a $p$-value =
  1 (0.1). (Right) Regions in $M_2$-$\mu$ plane corresponding to
  various values of $d_n/d_e$ in split SUSY, obtained by varying
  $g_T^{u,d,s}$ within our estimated uncertainties.  In the bands of
  constant $d_n/d_e$, the values of both $d_n$ and $d_e$ decrease as
  $\mu$ and $M_2$ increase. Using $d_e \le 1.1 \times 10^{-29} $ e
  cm~\protect\cite{Andreev:2018ayy} and assuming maximal $CP$ violation
  ($\sin \phi = 1$), the allowed region lies above the solid black
  line. For $\mu, M_2 > 500$~GeV, maximizing the ratio $d_n/d_e$ along
  this line gives the upper bound $d_n < 4.1 \times 10^{-29}$ e cm at 
  $d_n/d_e=3.71$.
\label{fig:nEDM}}
\end{figure*}

\section{Comparison with Previous Work}
\label{sec:comparison}

In Table~\ref{tab:resultsFINAL}, we show that results obtained by the
ETMC collaboration~\cite{Alexandrou:2017qyt} using a single physical
mass ensemble generated with 2-flavors of maximally twisted mass
fermions with a clover term at $a=0.0938(4)$~fm, $M_\pi=130.5(4)$~MeV
and at much smaller $M_\pi L = 2.98$ agree with our more complete 
analysis. Such consistency is expected if the differences due to the
number of dynamical flavors, and possible discretization and finite-volume 
corrections in the ETMC'17 results are small or
cancel. \looseness-1

\section{Implications for neutron electric dipole moment}
\label{sec:nEDM}

The tensor charges for the neutron are, in the isospin symmetric
limit, obtained from the proton charges by interchanging the 
light-quark labels, $u \leftrightarrow d$. Using the values given in
Table~\ref{tab:resultsFINAL} and the experimental bound on the nEDM
($d_n \le 2.9\times 10^{-26}\ e$ cm~\cite{Baker:2006ts}), the
relation
\begin{equation}
d_n = d_u^\gamma g_T^u + d_d^\gamma g_T^d + d_s^\gamma g_T^s \,,
\label{eq:dn}
\end{equation}
provides constraints on the $CP$ violating quark EDMs, $d_q^\gamma$, arising in
BSM theories, assuming that the quark EDM is the only $CP$-violating BSM
operator. The bounds on $d_q^\gamma$ are shown in the left panel of
Fig.~\ref{fig:nEDM}.  Of particular importance is the reduction in the
error in $g_T^s$ compared to our previous result, $g_T^s=0.008(9)$, in
Ref.~\cite{Bhattacharya:2015esa}.  The new result lets us bound
$d_s^\gamma$. Conversely, the overall error in $d_n$ is reduced even
if $d_s^\gamma$ is enhanced versus $d_u^\gamma$ by $m_s/m_u \approx
40$ as occurs in models in which the chirality flip is provided by the
Standard Model Yukawa couplings.

In general, BSM theories generate a variety of $CP$-violating operators
that all contribute to $d_n$ with relations analogous to
Eq.~\eqref{eq:dn}. As discussed in Ref.~\cite{Bhattacharya:2015esa},
in the ``split SUSY''
model~\cite{ArkaniHamed:2004fb,Giudice:2004tc,ArkaniHamed:2004yi}, the
fermion EDM operators provide the dominant BSM source of $CP$ violation.
In Fig.~\ref{fig:nEDM} (right), we therefore update the contour plots for $d_n/d_e$ in the
gaugino ($M_2$) and Higgsino ($\mu$) mass parameter plane with the
range $500$~GeV to $10$~TeV. For this analysis, we have followed
Ref.~\cite{Giudice:2005rz} and set $\tan \beta = 1$. 

Thanks to the greatly reduced uncertainty in the tensor charges
(factor of $\approx 6$ for $g_T^s$ and $\approx 2$ for $g_T^l$), the
ratio $d_n/d_e$ is much more precisely known in terms of SUSY mass
parameters.  This allows for stringent tests of the split SUSY
scenario with gaugino mass
unification~\cite{ArkaniHamed:2004fb,Giudice:2004tc,ArkaniHamed:2004yi}.
In particular, our results and the experimental bound $d_e < 1.1
\times 10^{-29} e$~cm~\cite{Andreev:2018ayy,Baron:2013eja}, imply the split-SUSY upper
bound $d_n < 4.1 \times 10^{-29} e$~cm. This limit is falsifiable by the
next-generation nEDM experiments. Constraints on split SUSY from LHC
searches predicated on gluino decays rule out the region below about a
TeV in the $\{ \mu,M_2 \}$ plane~\cite{Sirunyan:2018vjp}, whereas,
assuming a maximal $CP$-violating phase ($\sin \phi = 1$), EDMs currently
probe scales considerably higher than LHC's reach.\looseness-1

\section{Conclusions}
\label{sec:conclusions}

We present results for the flavor diagonal tensor charges, $g_T^{u}$,
$g_T^{d}$ and $g_T^{s}$, with control over all the systematics for
both the connected and the disconnected contributions.  The light
disconnected contributions, which were neglected in the PNDME'15
publication~\cite{Bhattacharya:2015wna}, are small and show little
variation versus the lattice spacing or the pion mass.  The errors in
the individual connected and the disconnected contributions on each
ensemble have been significantly reduced due to the high-statistics.
The final results, given in Table~\ref{tab:resultsFINAL}, were
obtained using a controlled chiral-continuum fit to data on multiple
ensembles that cover a sufficiently large range in lattice spacing and
pion mass. The reduced errors have allowed us to tighten the
constraints on the quark EDM couplings and on the ratio $d_n/d_e$ in
the split SUSY scenario with gaugino mass
unification~\cite{ArkaniHamed:2004fb,Giudice:2004tc,ArkaniHamed:2004yi}
as shown in Fig.~\ref{fig:nEDM}.

\begin{acknowledgments}
We thank the MILC Collaboration for providing the 2+1+1-flavor HISQ
lattices used in our calculations. The calculations used the Chroma
software suite~\cite{Edwards:2004sx}. Simulations were carried out on
computer facilities of (i) the National Energy Research Scientific
Computing Center, a DOE Office of Science User Facility supported by
the Office of Science of the U.S. Department of Energy under Contract
No. DE-AC02-05CH11231; and, (ii) the Oak Ridge Leadership Computing
Facility at the Oak Ridge National Laboratory, which is supported by
the Office of Science of the U.S. Department of Energy under Contract
No. DE-AC05-00OR22725; (iii) the USQCD Collaboration, which are funded
by the Office of Science of the U.S. Department of Energy, and (iv)
Institutional Computing at Los Alamos National Laboratory.
T. Bhattacharya and R. Gupta were partly supported by the
U.S. Department of Energy, Office of Science, Office of High Energy
Physics under Contract No.~DE-AC52-06NA25396.  T. Bhattacharya,
V. Cirigliano, R. Gupta, Y-C. Jang and B. Yoon were partly supported by
the LANL LDRD program.  The work of H.-W. Lin is supported by the US National
Science Foundation under grant PHY 1653405 ``CAREER: Constraining
Parton Distribution Functions for New-Physics Searches''.
\end{acknowledgments}

\clearpage
%
\bibliography{ref} 

\end{document}